\documentclass[prb,amsmath,amssymb,showpacs,twocolumn]{revtex4}
\usepackage{graphicx}
\def\be{\begin{equation}}
\def\ee{\end{equation}}
\def\om{\omega}

\def\bo{B_{\rm o}}
\def\bi{B_{\rm i}}

\def\e{e^*}
\def\m{m^*}
\def\FI{\omega_{\Phi}}
\def\X{\omega_{\xi}}
\def\K{\omega_{\kappa}}

\begin{document}
\title{Nonlinearity of the field induced by a rotating superconducting shell}
\author{Jorge Berger}
\affiliation{Physics Unit, Ort Braude College, P. O. Box 78,
21982 Karmiel, Israel and \\
Department of Physics, Technion, 32000 Haifa, Israel}
\email{phr76jb@tx.technion.ac.il}
\begin{abstract}
For a thin superconducting shell with cylindrical symmetry, the magnetic field
generated by its rotation is easily evaluated in the Ginzburg--Landau framework.
We compare this field with the result that is obtained by using the London theory.
\end{abstract}
\pacs{74.20.De}%
\maketitle

\section{INTRODUCTION}
Almost a century ago Barnett\cite{Barnett} found that ``any magnetic substance
becomes magnetized when set into rotation" ``by a sort of molecular gyroscopic
action." In 1933 Becker {\it et al.}\cite{Becker} found that the magnetic field
induced by a ``perfect conductor" is 
\be
B=-(2\m c/\e)\omega \;,
\label{Bec}
\ee
where $\m$ and $\e$ are the mass and charge of a free charge carrier, $c$ is the
speed of light and $\omega$ is the angular velocity at which the body rotates.
This is precisely the field that according to Barnett would be generated by a
``perfectly diamagnetic" material.
Since all the vectors in this article will be directed along the axis of rotation,
we will disregard their vectorial character.

The ``perfect conductor" presents a conceptual difficulty, since the currents in it
depend on the initial conditions. On the the other hand, for a superconductor, a 
thermodynamically significant result can be obtained. Using the London theory\cite{London}
and in the absense of an external field, Eq.~(\ref{Bec}) is recovered. $\m$ and 
$\e=-2e$ are now the mass and charge of a Cooper pair. 
The rotating superconductor was analyzed in the Ginzburg--Landau (GL) framework by
Verkin and Kulik;\cite{Ku2} this kind of analysis has been recently revised by Capellmann.\cite{Cap}

Equation (\ref{Bec}) has been verified by several experiments,
\cite{exp1,exp15,exp2,exp3,exp4,exp5} and they have actually become a means for measuring
the effective mass of a Cooper pair. The possibility of measuring the charge to mass ratio from the magnetic field generated by a rotating superconductor has been regarded as an example that stands on equal footing with ``quantum protection" and symmetries, i.e., a case in which the result is insensitive to microscopics.\cite{Laugh} Denoting by $m$ the mass of an electron, the most precise result obtained\cite{exp4} to my knowledge is $\m/2m=1.000084(21)$. The relationship
between the effective and the bare mass of the charge carriers has been analyzed
from the microscopic\cite{M1,M2,M3,M4} and the thermodynamic\cite{T1,T2} points of
view, and none of the results is in full agreement with experiment.

The possibility of an electric field generated by a rotating\cite{El} or even
static\cite{Hirsch} superconductor has also been considered.

In spite of the fundamental character that has been attributed to Eq.~(\ref{Bec}), this article raises the point that it is just an approximation, valid
for low angular velocities. For high rotation speeds $B$ will not be a linear---and
not even a monotonic---function of $\omega$. This will be shown in the following
section, by considering a superconducting sample with a shape such that the 
GL theory is very easily applied.

\section{OUR MODEL}
We consider a long superconducting cylindrical shell with radius $R$ and thickness $d$ 
($d\ll R$) that rotates around its axis with angular velocity $\om$. 
The entire analysis will be conducted from an inertial frame of reference.
Let $\bo$ (resp. $\bi$) 
denote the component of the magnetic field in the direction of the axis outside 
(resp. inside) the cylinder. ($\bo$ and $\bi$ are assumed to be uniform.) 
Let $N$ be the number of superconducting pairs per unit area and $v'$ 
their velocity relative to the ions of the shell. 
It follows that the current per unit length is $N\e v'$ and, by Amp\`{e}re's law,
the inner and outer fields are related by 
\be
\bi=\bo+\frac{4\pi}{c}N\e v' \;.
\label{Amp}
\ee
Quantization of the canonical momentum requires
\be
2\pi R\m v+\frac{\e}{c}\pi R^2\bi=Lh
\label{quant}
\ee
where $v$ is the velocity of the pairs relative to 
the laboratory, $L$ is an integer that determines the trapped flux and $h$ is Planck's 
constant. $v$ and $v'$ are related by
\be
v=v'+\om R
\label{rel}
\ee

Solving the system of equations (\ref{Amp})--(\ref{rel}) we obtain
\begin{eqnarray}
\bi-\bo&=&\frac{2\m c\gamma(\FI-\om)}{\e(1+\gamma)} \label{bi} \\
v&=&R(\gamma\om+\FI)/(1+\gamma) \label{v}
\end{eqnarray}
where we have defined
\be
\gamma=\frac{2\pi(\e)^2 RN}{\m c^2} \; ; \;\;\;
\FI=\frac{Lh}{2\pi R^2\m}-\frac{\e\bo}{2\m c}
\label{defs}
\ee
Expressions analogous to (\ref{bi}) have been found in 
Refs.~[\onlinecite{Ku2,Ku1}].

If $\gamma\gg 1$ (for $R\sim 1$cm this means $N\gg 10^{12}{\rm cm}^{-2}$),
Eq.~(\ref{bi}) reduces to
\be
\bi-\bo=-\frac{2\m c}{\e}(\om-\FI)
\label{Lond}
\ee
which for $L=0$ and in the absense of external field is just Becker's result. 

The condition $\gamma\gg 1$ is easily fulfilled, and this seems to be the reason
that Eq.~(\ref{Bec}) is so widely accepted. In London's theory, $N$ is effectively
a constant. Even in Ref.~\onlinecite{Cap}, where $N$ could in principle be
obtained from the GL theory, in practice $\bi$ was evaluated
only in the case that ``stiffness of the wave function" can be assumed.
However, if the shell is very thin and the temperature close to critical, $N$
can be noticeably dependent on $\om$ and this dependence will be the source
of nonlinearity of $\bi(\om)$.

The density of pairs $N$ is obtained by minimizing the free energy. If the
thickness $d$ of the shell is small compared with the coherence length $\xi$,
the free energy per unit length is
\begin{eqnarray}
G&=&2\pi RN\left(\frac{1}{2}\m v^2+\alpha+\frac{\beta}{2d}N\right)
+\frac{R^2}{8}(\bi-\bo)^2 \nonumber \\
&-&\pi R(N_{\rm T}-2N)m(\om R)^2
\label{G}
\end{eqnarray}
where $\alpha$ and $\beta$ are the GL coefficients and $N_{\rm T}$
is the total number of electrons per unit area of the shell. The first term
consists of the kinetic and the condensation energy of the pairs and the second
is the contribution of $\bi$ to the free energy for given $\bo$. The last term
is due to the normal electrons, which have density $N_{\rm T}-2N$ and energy
per unit length $\pi R(N_{\rm T}-2N)m(\om R)^2$. Noting that\cite{Cap} $\om$ acts
as a Lagrange multiplier of the angular momentum of the normal electrons, we obtain 
the opposite sign.

Substituting the expressions (\ref{bi}) and (\ref{v}), defining the frequencies
\begin{eqnarray}
\X&=&\left(\frac{-2\alpha}{\m R^2}\right)^{1/2}=\frac{\hbar}{\m R\xi} \nonumber \\
\K&=&\left(\frac{\beta c^2}{2\pi(\e)^2 dR^3}\right)^{1/2}=\frac{\kappa\hbar}
{\m d^{1/2}R^{3/2}}
\label{sat}
\end{eqnarray}
where $\kappa$ is the GL parameter, and droping the constant $N_{\rm T}$, the free
energy can be rewritten as
\be
G=\left(\frac{\m cR}{\e}\right)^2\frac{\gamma}{2}\left(\gamma\K^2-\X^2+
\frac{\FI^2+\gamma\om^2}{1+\gamma}+\frac{2m}{\m}\om^2\right) \;.
\label{newG}
\ee

\begin{figure}[t]
\scalebox{0.85}{\includegraphics{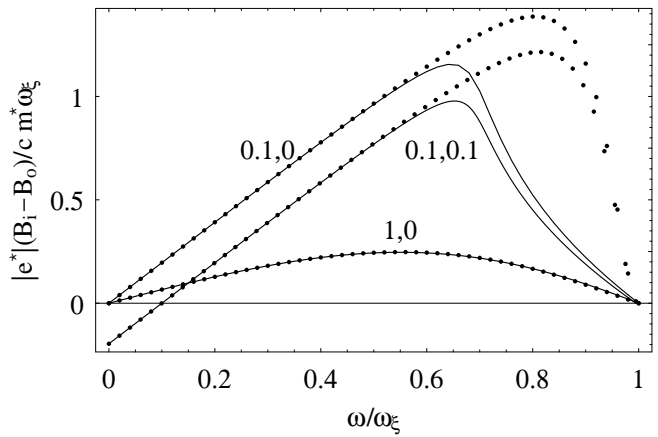}}%
\caption{\label{res}Magnetic field generated by the rotating shell as a function of its
angular velocity. Each curve is marked by the parameters $(\K/\X,\FI/\X)$. The effective
mass of a pair was taken as exactly equal to the mass of two electrons. The curves were
obtained by minimization of $G$ in Eq.~(\ref{newG}) and the dots from the
approximation (\ref{app}). From Eq.~(\ref{Bec}) one would obtain a straight line 
through the origin with slope 2.}
\end{figure}

\section{RESULTS}
Let us first assume that the fluxoid number $L$ stays unchanged.
For given $\om$, $\bo$, $L$, and $\m/m$, $N$ can be obtained by numerical minimization of $G$
in Eq.~(\ref{newG}). A fair approximation, that fits $N$ to order $O(\FI^2,\m-2m)$ at $\om=\X$ and to order $O(\om^4,\om^2(\m-2m),\FI^4)$ for $\om\ll\X$ is
\be
\gamma=\frac{\X^2-\om^2}{2\K^2}\left(1-\frac{4\FI^2\K^4/\X^2+\om^2(\X^2+4\K^2)}
{(\X^2+2\K^2)^2}\right) \;.
\label{app}
\ee
Representative results are shown in Fig.~\ref{res}. 

Let us now revise the assumption that $L$ stays unchanged. $L$ is related to
$\FI$ by definition (\ref{defs}). If $L$ could change continuously, $\FI$ would
assume the value that minimizes $G$ in Eq.~(\ref{newG}), i.e., $\FI=0$. Since
$L$ has to be integer, the most stable value of $L$ will be the one that gives
the lowest possible value of $\FI$. Experiment~[\onlinecite{exp4}] shows that
several values of $L$ (within the range of about five units from the one of
maximal stability) are also possible; they give rise to metastable states.
In order to evaluate the potential barriers between different values of $L$ we
would have to consider situations that deviate from cylindrical symmetry; however,
the form of Eq.~(\ref{newG}) suggests that there is no coupling between $\om$ and
$\FI$ and a state that is metastable for the sample at rest will remain metastable
when the sample is rotating and fluxoid jumps will be rare events. The conditions for metastability of different fluxoid states were studied in detail in Ref.~\onlinecite{Ku2}.

\section{DISCUSSION}
We have found that the magnetic field generated by a rotating superconducting shell
is not a linear function of the angular velocity; rather, it reaches a maximum for
angular velocities of the order of $\X$. $\X$ is a decreasing function of the coherence
length and is smallest close to the critical temperature. For $R\sim 1$cm and 
$\xi\sim 10^{-4}$cm, $\X\sim 10^4{\rm s}^{-1}$. Even if the angular velocity is
smaller than $\X$ by a few orders of magnitude, nonlinearity ought to be taken into
account for the purpose of precision measurements.

We have only considered the case of a very thin shell, with perfect cylindric
symmetry. The thinner the shell, the larger the value of $\K$ and Fig.~\ref{res}
shows than this implies an initial slope that is smaller than in Becker's result.
However, nonlinearity shows up also when the shell is less thin, suggesting that
this qualitative feature will be present for every rotating superconductor, regardless
of its shape.

The induced magnetic field depends also on $\FI$. $\FI$ depends on the mismatch between
the applied magnetic flux and the angular momentum per Cooper pair that would lead to
a vanishing velocity. This mismatch is usually of the order of a quantum. In this case
we have $\FI/\X\sim\xi/R$. Since this ratio is usually very small, the $\FI$-dependence
is expected to be a minor effect when regarded in the entire scale $-\X\le\om\le\X$.
Jumps between different possible values of $\FI$ are expected to be rare, but when $\om$
approaches $\X$ the density of superconducting pairs becomes small and the potential
barriers between different states have to be small, too. In addition, if the thickness
of the shell is not exactly uniform, continous passage between different states may
become possible.\cite{vortex} 

\begin{acknowledgments}
This work has been supported in part by the Israel Science Foundation.
\end{acknowledgments}

\end{document}